\documentclass[
superscriptaddress,
 amsmath,amssymb,
 aps,prb, two column
]{revtex4-1}

\usepackage{epsf,verbatim,graphicx,caption,dsfont,tikz-feynhand}
\usepackage{dcolumn,subfigure}
\usepackage{bm,bbm}

\newcommand{\mi}{\mathbbm{i}}  
\newcommand{\ket}[1]{\left| #1 \right>} 

	\newcommand{\mean}[1]{\langle #1 \rangle} 
	\newcommand{\elem}[3]{\langle #1|#2|#3\rangle}
	\newcommand{\beq}{\begin{eqnarray}}
	\newcommand{\eeq}{\end{eqnarray}}

\begin{document}


\title{Conductivity in nodal line semimetals with short-ranged impurity potentials}

\author{Hui Yang}
\affiliation{International Center for Quantum Materials, School of Physics, Peking University, Beijing 100871, China
}
\author{Fa Wang}
\affiliation{International Center for Quantum Materials, School of Physics, Peking University, Beijing 100871, China
}
\affiliation{Collaborative Innovation Center of Quantum Matter, Beijing 100871, China
}

\date{\today}

\begin{abstract}
We study the transport properties in nodal line semimetals with short-ranged impurity potentials at zero temperature. By computing the Drude conductivity and the corrections from the interference of particle and hole trajectories,  we find that the electrons are localized in directions both parallel and perpendicular to the plane of nodal ring. We further calculate the conductivity in a weak magnetic field, and find that the perpendicular magnetic field totally destroys the weak localization gives a positive quantum interference correction, which is similar to the result in Weyl semimetals. But for a parallel magnetic field, because of the $\pi$ Berry phase of the electron orbit around the nodal line, the magnetoconductivity is negatively proportional to $B$. The difference between the perpendicular and the parallel magnetic field may be verified by experiments. Nodal line semimetals which break inversion and time-reversal symmetry and have spin-orbit coupling are also considered and produce qualitatively the same results.

\end{abstract}

\maketitle


\section{Introduction}
Topological materials\cite{RevModPhys.82.3045, RevModPhys.83.1057, RevModPhys.90.015001} have attracted much interests in both theoretical and experimental aspects in the past decades, including both gapped topological phases and gapless phases. The Weyl semimetal\cite{wan_topological_2011, PhysRevX.5.031013, Xu613} is an example of gapless materials with surface Fermi arcs, which may give rise to interesting magneto-response, such as the quantum oscillation\cite{pereg-barnea_quantum_2009, zhang_quantum_2016, potter_quantum_2014} from the surface state or the quantum Hall effect in 3D\cite{Zhang2018, PhysRevLett.119.136806}.
The idea of nodal line semimetals\cite{PhysRevB.84.235126, PhysRevLett.116.127202, PhysRevB.92.081201, PhysRevB.90.115111, PhysRevB.92.045108, PhysRevLett.115.026403} is a generalization of Weyl semimetals. In nodal line semimetals, the valence band and the conduction band touch at a closed loop\cite{PhysRevLett.116.127202, PhysRevLett.116.195501, PhysRevB.93.020506} which possesses some non-trivial topology. Like the other symmetry protected topological phases\cite{doi:10.1146/annurev-conmatphys-031214-014740, Gu2014, Wen2013, Chen2013, PhysRevB.85.085103}, the nodal line is robust against perturbations which preserve the symmetry\cite{doi:10.1063/1.3149495}. The nodal line can either be gapped or evolve into other nodal semimetals such as Dirac or Weyl semimetals when the symmetry is broken\cite{PhysRevB.92.081201}. There are many theoretical proposals for the realization of nodal line semimetals both with\cite{PhysRevLett.115.036806, PhysRevLett.116.127202} or without\cite{PhysRevLett.117.096401, PhysRevB.96.115201, PhysRevLett.116.195501} spin-orbit coupling. Many experiments have realized the nodal line materials\cite{wu_dirac_2016, PhysRevB.93.201104, PhysRevLett.117.016602}, such as PtSn$_4$, ZrSiSe and ZrSiTe.\\

One interesting feature of the nodal line semimetal is the non-trivial $\pi$ Berry phase, which can be extracted from quantum oscillation. The signature of quantum oscillation\cite{Ashcroft} in nodal line semimetal has been studied both theoretically\cite{PhysRevB.97.165118, PhysRevLett.120.146602, PhysRevB.97.205107, PhysRevB.94.195123} and experimentally\cite{PhysRevLett.117.016602}. The response to the magnetic field depends on the geometry of the Fermi surface. For large chemical potential, the Fermi surface is in the shape of drum, which gives a topological trivial Berry phase. For small chemical potential, the Fermi surface is a torus. In this case the electrons  will acquire a non-trivial Berry phase when the extreme orbital encircling around the nodal ring, which is related to the pseudospin texture\cite{PhysRevLett.118.016401} and can be extracted from the Landau-fan diagram\cite{shoenberg_2009}.\\

Besides, the non-trivial Berry phase also has impact on magneto-transport properties. The experiment in reference\cite{Alie1601742} show that the magnetoresistivity $\delta\rho=c_1H+c_2H^2$. When $H\parallel a$, $c_2=0$, which means $\delta\rho$ is proportional to the strength of magnetic field. Motivated by this observation, in this paper, we study the electronic transport properties in nodal line semimetals with short-ranged potential, including $\delta$ function potential and the screened Coulomb potential. We systematically calculate the Drude conductivity and the quantum interference corrections. We find the quantum interference correction in nodal line semimetal is negative in both $x$ and $z$ direction, which is a signature of weak localization. Under a weak magnetic field, the quantum interferences follow different field dependence for perpendicular magnetic field and parallel magnetic field because of the $\pi$ Berry phase. For the case of perpendicular magnetic field, the weak-localization is destroyed, which give a $\delta\sigma^{qi}(B)\propto\sqrt{B}$ or $\propto B^2$ dependence in different limits. For a parallel magnetic field, the field dependence is $\delta\sigma^{qi}(B)\propto-B$. \\

The paper is organized as follows. In Sec. II, we introduce the nodal line Hamiltonians we are going to study in this paper and the scattering potential. In Sec. III, we calculate the Drude conductivity, and quantum interference. In Sect. IV, we find the field dependence in the presence of weak magnetic field. We end and give a conclusion in Sec. V. The detailed derivations are presented in the Appendix.

\section{The nodal line Hamiltonian}
The nodal line Hamiltonian is given by\cite{PhysRevLett.115.036806}
\beq
H(k_x, k_y, k_z)=\frac{1}{2}(\frac{k_x^2}{m_x}+\frac{k_y^2}{m_y}-\frac{k_0^2}{m_0})\sigma_x+v_zk_z\sigma_y,
\eeq
in which the $\sigma$'s are pseudospin operator acting on the orbital space and $v_z$ is the dispersion in the $k_z$ direction. We have chose the parameters $m_x=m_y=m_0=1$. This Hamiltonian preserves time-reversal and inversion symmetry and it does not include spin-orbit coupling. The spectrum is given by $E_{\pm}=\pm\sqrt{\left(\frac{1}{2}(k_x^2+k_y^2-k_0^2)\right)^2+v_z^2k_z^2}$. The conduction band and valence band touch at a circle in momentum space, which is given by $k_x^2+k_y^2=k_0^2$ and $k_z=0$, and it is protected by the mirror symmetry, which acts on the Bloch Hamiltonian $H({\bf k})={\bf h}({\bf k})\cdot{\bf \sigma}$ as ${\bf h}({\bf k})\rightarrow(-h_x(k_x,k_y,-k_z), -h_y(k_x,k_y,-k_z), h_z(k_x,k_y,-k_z))$. So there may be a nodal ring living on the mirror plane $k_z=0$ with $h_x=h_y=0$. The topology of the Fermi surface will be different for increasing chemical potential. It evolves to a drum with genus $0$ from a torus with genus $1$ which enclose a degenerate line. In the following, we will focus on the case of small chemical potential $0\leq\mu\ll k_0^2$. The eigenstate of the conduction band $E_+$ is
\beq
\ket{\bf k}=\frac{1}{\sqrt{2}}(1, e^{\mi\theta})^{\mathrm{T}},
\eeq
where we have introduced $\theta$, $\phi$ to parameterize ${\bf k}$ as $k_x=\sqrt{2E\cos\theta+k_0^2}\cos\phi$, $k_y=\sqrt{2E\cos\theta+k_0^2}\sin\phi$ and $k_z=\frac{E}{v_z}\sin\theta$. Under inversion ${\bf k}\rightarrow-{\bf k}$, $\theta$ and $\phi$ transform as $(\theta,\phi)\rightarrow(-\theta,\phi+\pi)$. From the eigenstate given in Eq.(2), we find there will be a $\pi$ Berry phase when the orbit encircles the nodal ring. The origin of the $\pi$ Berry phase can also be understand as follows. Expanding the Hamiltonian around the nodal ring $k_z^2+k_y^2=0$ and $k_z=0$, then we will arrive at a Dirac Hamiltonian, and the orbit winding around the Dirac point will acquire a $\pi$ Berry phase.\\

\begin{figure}
\begin{center}
\includegraphics[width=8.9cm]{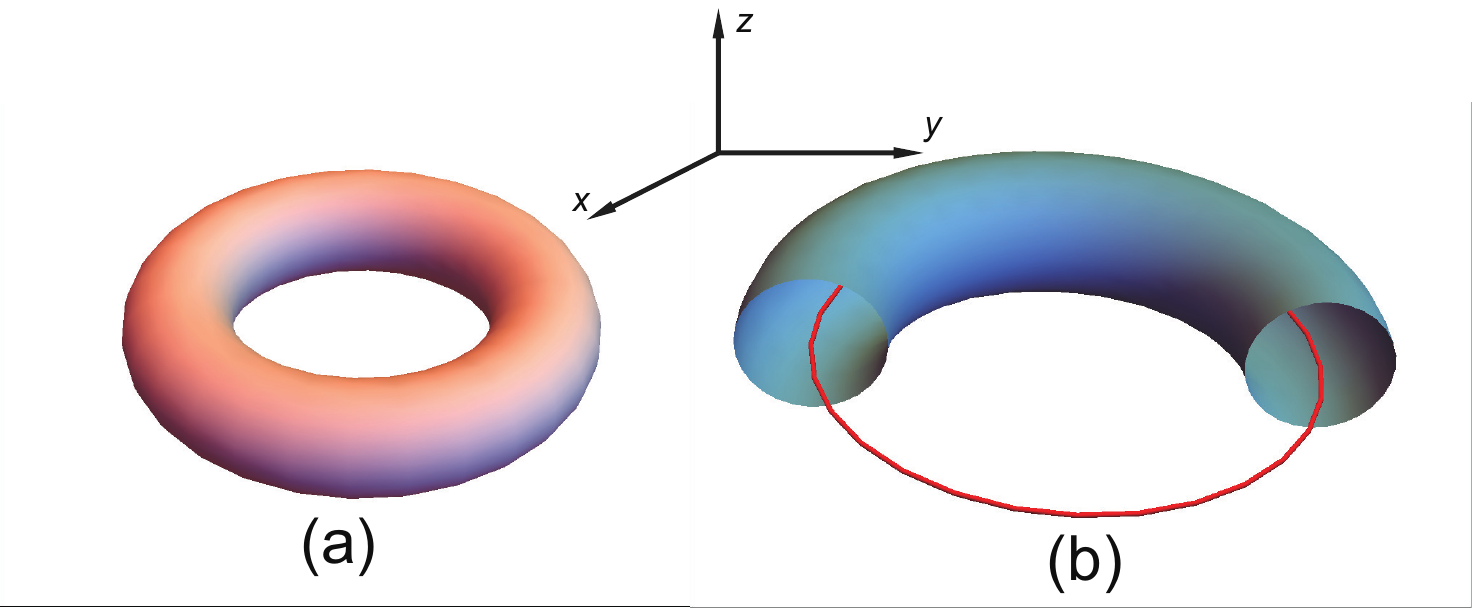}
\end{center}
\caption{Fermi surface (a) and the nodal line (b) of Hamiltonian in Eq.(1). }
\end{figure}

We assume the scattering potential is isotropic and short-ranged. In the following calculation, first we approximate it as a $\delta$ function potential $U({\bf r})=u\sum_i\delta({\bf r}-{\bf R_i})$, where $u$ is the strength of the scattering potential. We assume the spatial correlation of the $\delta$ potential is $\mean{U(r)U(r^\prime)}\propto\delta(r-r^\prime)$. Then we will consider the case of Yukawa potential, which is given by $\frac{1}{4\pi r}e^{-\lambda r}$.

We will compare the result of Hamiltonian (1) with the case with spin-orbit coupling and with the case for breaking inversion and time-reversal symmetry. Under time reversal and inversion, the Bloch Hamiltonian transforms as $H({\bf k})\rightarrow H^*(-{\bf k})$, $H({\bf k})\rightarrow \sigma_x H(-{\bf k})\sigma_x$, respectively.
The Hamiltonian with spin-orbit coupling is give by\cite{PhysRevB.92.081201}
\beq
H(k_x,k_y,k_z)=k_x\tau_0s_x+k_y\tau_2s_y+k_z\tau_0s_z+\Delta\tau_xs_x,
\eeq
which will give the same nodal line with Eq.(1). $\tau_1$ and $\sigma_i$ are Pauli matrices acting on two isospin spaces. The nodal ring will shrink to a point from $m>0$ to $m=0$ and then it will increase from $m=0$ to $m<0$ but now the nodal ring is give by the other two band.

The Hamiltonian without time reversal and inversion symmetry is given by 
\beq
H(k_x,k_y,k_z)\!=\!\left(\frac{1}{2}(k_x^2\!+\!k_y^2\!-\!k_0^2)\!+\!\lambda vk_z\right)\sigma_x\!+\!vk_z\sigma_y,
\eeq
where the term $k_z\sigma_x$ breaks time-reversal and inversion symmetry.

\section{Calculation of the Drude conductivity and the quantum interference}
\subsection{Drude conductivity}
The semiclassical Drude conductivity is given by the formula,
\beq
\sigma_{ii}^{sc}=\frac{e^2}{2\pi}\sum_k\tilde{v}_iv_iG_k^RG_k^A,
\eeq
where $v_i$ is the bare velocity along $i$ direction, which is given by $\elem{{\bf k}}{\frac{\partial H}{\partial k}}{{\bf k}}$, and $\tilde{v}_i$ is the velocity corrected by the scattering potential, and $G_k^{R(A)}$ is the retarded (advanced) Green function.

 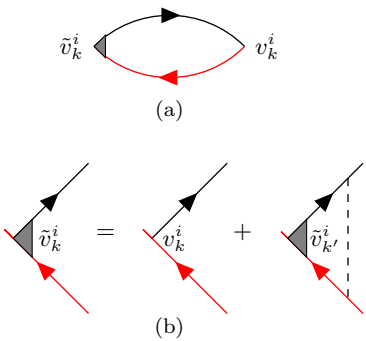
\begin{figure}[htbp]
 \centering
 \subfigure[]
 {\begin{minipage}[t]{0.2\textwidth}
\centering
\begin{tikzpicture}
\begin{feynhand}
\vertex (a1) at (0,0); \vertex (a2) at (2,0);
\propag [fer, black,in=135, out=45,line width=2mm] (a1) to  (a2);
\propag [fer, red,in=315, out=225, line width=2mm] (a2) to (a1);
\node at (-0.3,0) {$\tilde{v}_k^i$};\node at (2.3,0) {${v}_k^i$};
\filldraw[fill=gray] (0,0) -- (1.5mm,-1.5mm) to (1.5mm,1.5mm) -- cycle;
\end{feynhand}
\end{tikzpicture}
\end{minipage}}
\subfigure[]{
 \begin{minipage}[t]{0.2\textwidth}
\centering
\begin{align*}
\begin{tikzpicture}[baseline=(a1.base)]
\begin{feynhand}
\vertex (a1) at (0,0); \vertex (a2) at (1,1);\vertex(a3) at (1,-1);
\propag [fer, black,in=225, out=45,line width=2mm] (a1) to  (a2);
\propag [fer, red,in=135, out=135, line width=2mm] (a3) to (a1);
\filldraw[fill=gray] (0,0) -- (0.25,-0.25) to (0.25,0.25) -- cycle;
\node at (0.5,0) {$\tilde{v}_k^i$};
\end{feynhand}
\end{tikzpicture}
=
\begin{tikzpicture}[baseline=(a1.base)]
\begin{feynhand}
\vertex (a1) at (0,0); \vertex (a2) at (1,1);\vertex(a3) at (1,-1);
\propag [fer, black,in=225, out=45,line width=2mm] (a1) to  (a2);
\propag [fer, red,in=135, out=135, line width=2mm] (a3) to (a1);
\node at (0.3,0) {${v}_k^i$};
\end{feynhand}
\end{tikzpicture}
+
\begin{tikzpicture}[baseline=(a1.base)]
\begin{feynhand}
\vertex (a1) at (0,0); \vertex (a2) at (1,1);\vertex(a3) at (1,-1);
\propag [fer, black,in=225, out=45,line width=2mm] (a1) to  (a2);
\propag [fer, red,in=135, out=135, line width=2mm] (a3) to (a1);
\filldraw[fill=gray] (0,0) to (0.25,-0.25) to (0.25,0.25) -- cycle;
\node at (0.5,0) {$\tilde{v}_{k^\prime}^i$};\propag[sca] (0.8,0.8) to (0.8,-0.8);
\end{feynhand}
\end{tikzpicture}
\end{align*}
\end{minipage}}
\caption{Drude conductivity (a) and velocity (b).}
\end{figure}
In the Born approximation, the scattering rate $\tau_k$ is give by
\beq
\frac{1}{\tau_k}=2\pi\sum_{k^\prime}\mean{|U_{k^\prime,k}|^2}_{dis}\delta(E_F-\epsilon_{k^\prime})=\pi N_Fnu^2,
\eeq
where $\mean{|U_{k_1,k_2}|^2}_{dis}=\frac{nu^2}{2}(1+\cos(\theta_1-\theta_2))$ with $U_{k,k^\prime}=\elem{k}{U}{k^\prime}$ and $n$ is the impurity density. We can get the Drude conductivity from Eq.(5),
\beq
\sigma_{xx}^{sc}=\sigma_{yy}^{sc}=\frac{e^2N_Fk_0^2\tau}{4},\sigma_{zz}^{sc}=e^2N_Fv_z^2\tau,
\eeq
from which we can define the diffusion coefficient $D_1=\frac{k_0^2\tau}{4}$ and $D_2=v_z^2\tau$.

\subsection{Quantum interference}
In the quantum diffusion regime, the particle and hole trajectories will interfere with each other. The quantum interference will contribute to the classical conductivity, which may lead to weak localization or weak anti-localization, depending on the relative phase of the two paths. The Feynman diagrams of the quantum interference is given in Fig. 2.

 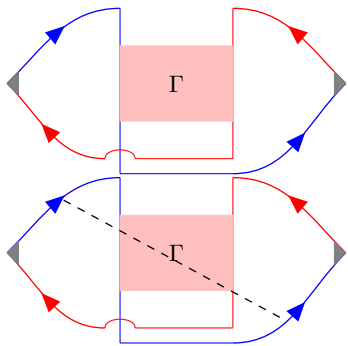
\begin{figure}[htbp]
\centering
\begin{tikzpicture}
\begin{feynhand}
\vertex (a1) at (0,0); \vertex (a2) at (1.5,1);\vertex (a3) at (3,1);\vertex (a4) at (4.5,0);\vertex (a5) at (3,-1);\vertex (a6) at (1.5, -1.2);\vertex (a7) at (1.3,-1);\vertex (a8) at (1.7,-1);\vertex (a9) at (3,-1.2);
\propag [fer, blue,in=180, out=45,line width=2mm] (a1) to  (a2);
\propag [fer, red,in=0, out=135, line width=2mm] (a4) to (a3);
\propag[fer,blue,in=225,out=0] (a9) to (a4);
\propag[fer,red,in=315,out=180] (a7) to (a1);
\draw [blue] (a2) to  (a6);
\draw [red,in=90,out=90] (a7) to (a8);
\draw[red] (a8) to (a5);\draw[red] (a5) to (a3);
\draw[blue] (a9) to (a6);
\filldraw[fill=pink] (1.5,0.5) [pink]to (3,0.5)[red] to (3,-0.5)[pink] to (1.5,-0.5) -- cycle;
\node at (2.25,0) {\textcolor{black}{$\Gamma$}};
\filldraw[fill=gray] (0,0) [gray] to (1.5mm,-1.5mm) to (1.5mm,1.5mm) -- cycle;
\filldraw[fill=gray] (4.5,0) [gray] to (4.35,-1.5mm) to (4.35,1.5mm) -- cycle;
\end{feynhand}
\end{tikzpicture}
\begin{tikzpicture}
\begin{feynhand}
\vertex (a1) at (0,0); \vertex (a2) at (1.5,1);\vertex (a3) at (3,1);\vertex (a4) at (4.5,0);\vertex (a5) at (3,-1);\vertex (a6) at (1.5, -1.2);\vertex (a7) at (1.3,-1);\vertex (a8) at (1.7,-1);\vertex (a9) at (3,-1.2);\vertex (a10) at (0.75,0.7);\vertex (a11) at (3.7,-0.9);
\propag [fer, blue,in=180, out=45,line width=2mm] (a1) to  (a2);
\propag [fer, red,in=0, out=135, line width=2mm] (a4) to (a3);
\propag[fer,blue,in=225,out=0] (a9) to (a4);
\propag[fer,red,in=315,out=180] (a7) to (a1);
\draw [blue] (a2) to  (a6);
\draw [red,in=90,out=90] (a7) to (a8);
\draw[red] (a8) to (a5);\draw[red] (a5) to (a3);
\draw[blue] (a9) to (a6);
\filldraw[fill=pink] (1.5,0.5) [pink]to (3,0.5)[red] to (3,-0.5)[pink] to (1.5,-0.5) -- cycle;
\node at (2.25,0) {\textcolor{black}{$\Gamma$}};
\propag[sca] (a10) to (a11);
\filldraw[fill=gray] (0,0) [gray] to (1.5mm,-1.5mm) to (1.5mm,1.5mm) -- cycle;
\filldraw[fill=gray] (4.5,0) [gray] to (4.35,-1.5mm) to (4.35,1.5mm) -- cycle;
\end{feynhand}
\end{tikzpicture}
\caption{The quantum interference to the conductivity type I (the upper diagram) and type II (the lower diagram).}
\end{figure}
The quantum interference correction is $\sigma^{qi}=\sigma_1^{qi}+2\sigma_2^{qi}$, where the factor 2 is because the number of the type II diagram in Fig. 3 is 2. First, we can calculate the scattering vertex, known as cooperon. In the static limit and long-wavelength limit, The cooperon of the nodal line semimetal is given by
\beq
\Gamma_q= \frac{1}{2\pi N_F\tau^2}\frac{1}{D_1(q_x^2+q_y^2)+2D_2q_z^2}.
\eeq
The quantum interference correction to the semiclassical conductivity is proportional to $\sum_q\Gamma_q$. We can get the results
\beq
\sigma_{xx}^{qi}=\sigma_{yy}^{qi}=-\frac{e^2}{2\pi^3}\sqrt{\frac{
D_1}{2D_2}}(\frac{1}{l_x}-\frac{1}{l_\phi})\propto\frac{k_0}{v_z},\nonumber\\
\sigma_{zz}^{qi}=-\frac{e^2}{2\pi^3}\sqrt{\frac{
D_2}{2D_1}}(\frac{1}{l_z}-\frac{1}{l_\phi})\propto\frac{v_z}{k_0}.
\eeq
here $l_\phi$ is the coherence length of electrons and we assume $l_\phi\ll l_i(i=x,y,z)$ and $l_i$ is the size of the sample in direction $i$, so the quantum interference is negative, which is a signature of weak localization of electrons.From Eq.(9), we see that for a give $v_z$ and $k_0$ there exits some critical value of the system size at which the quantum interference is isotropic.

\section{Conductivity in magnetic field}
In the presence of magnetic field, the momentum operator $\hat{\bf p}$ is replaced by $\hat{\bf p}-q{\bf A}$. Inserting it into the cooperon and transforming the cooperon into real space, we will find that $\Gamma(r)$ decay exponentially, which means the interference between particle and hole paths is destroyed by the magnetic field. This argument is valid for a perpendicular magnetic field, but it does not apply to a parallel magnetic field because of the $\pi$ Berry phase when the semiclassical orbital encircling the nodal ring.

In a magnetic field, the orbits are quantized as Landau levels with the quantization condition\cite{Ashcroft}
\beq
q_x^2+q_y^2=(n+\frac{1}{2}-\frac{\gamma}{2\pi})\frac{1}{l_B^2},
\eeq
where $\gamma$ is the Berry phase of the orbit and $l_B$ is the magnetic length $\sqrt{\hbar/4eDB}$. For perpendicular magnetic field, there is no Berry phase, so $q_x^2+q_y^2=(n+1/2)\frac{1}{l_B^2}$. But for parallel magnetic field, there is a $\pi$ Berry phase, so $q_y^2+q_z^2=n\frac{1}{l_B^2}$. Now the summation over ${\bf q}$ becomes summation over the Landau index and an integration about the direction of the magnetic field.\\

For perpendicular magnetic field, after scaling, the summation over q is\cite{PhysRevB.92.035203, PhysRevB.93.161110}
\beq
A_1&=&\sum_q\Gamma_q\propto\int dq_z\sum_n\frac{1}{(n+1/2)+l_B^2q_z^2}\nonumber\\&=&\int_0^\frac{1}{l_z}dq_z\left[\psi\left(\frac{l_B^2}{l_z^2}+l_B^2q_z^2+\frac{1}{2}\right)-\psi\left(\frac{l_B^2}{l_\phi^2}+l_B^2q_z^2+\frac{1}{2}\right)\right].\nonumber\\
\eeq
From Eq.(11) we can know the magnetoconductivity $\delta\sigma^{qi}(B)=\sigma(B)-\sigma(B=0)>0$, which means the weak localization is destroyed by the magnetic field, as expected before. In the limit of $l_\phi\gg l_B\gg l_z$, which can be approached at low temperature as said in Ref. [41], the magnetoconductivity $\delta\sigma^{qi}(B)\propto\sqrt{B}$. And in the limit of $l_B\gg l_\phi$ and $l_B\gg\l_z$, it is proportional to $B^2$. We find in this case the results are similar to the results for Weyl semimetals\cite{PhysRevB.92.035203}.\\

For the parallel magnetic field, Eq.(11) becomes
\beq
A_2&=&\sum_q\Gamma_q\propto\int dq_x\sum_n\frac{1}{n+l_B^2q_x^2}\nonumber\\&=&\int_0^\frac{1}{l_x}dq_x\left[\psi\left(\frac{l_B^2}{l_x^2}+l_B^2q_x^2\right)-\psi\left(\frac{l_B^2}{l_\phi^2}+l_B^2q_x^2\right)\right],\nonumber\\
\eeq
from which we see that the magnetoconductivity $\delta\sigma^{qi}(B)$ is negative and is always proportional to $B$.

From the analysis above, we have seen the effect of the Berry phase in nodal line semimetal, which leads to different magnetoconductivity and field dependence for the field direction parallel and perpendicular to the nodal ring plane.

\section{Conductivity with Yukawa potential}

We have calculated the transport properties with impurities of $\delta$ function potential before. Now we consider another type of short-ranged potential called Yukawa potential, which is given by 
\beq
U(r)=\frac{u}{4\pi\xi}\sum_i\frac{1}{|{\bf r}-{\bf R}_i|}e^{-|{\bf r}-{\bf R}_i|/\xi},
\eeq 
where $\xi$ corresponds to the interaction range of the potential. When $\xi=0$, it goes back to the case of $\delta$ function potential.
The scattering vertex is
\beq
\mean{U_{k_1,k_2}U_{k_2,k_1}}_{dis}=\mean{U_{k_1,k_2}U_{-k_1,-k_2}}_{dis}\nonumber\\=\frac{1}{(1+(\vec{k}_1-\vec{k}_2)^2\xi^2)^2}\frac{nu^2}{2}(1+\cos({\theta_1-\theta_2}).
\eeq
We calculate the scattering rate $1/\tau$ numerically for $\xi=0.6-1.5$ and the results are shown in Fig. 4, from which we can find the scattering rate   does not depend on the momentum approximately. We have $\frac{1}{\tau  nu^2}=0.082+(\text{k-dependent corrections})$ for $\xi=1$, $k_0^2=1$, $E_F=0.1$, and $v_z=0.8$ We find that there is no qualitatively difference when we calculate the velocity corrections and solve the vertex corrections numerically compared with the case of $\delta$ function potential. The vertex of this case is given by $\Gamma_q=\frac{1}{D_1^\prime(q_x^2+q_y^2)+D_2^\prime q_z^2}$.
\begin{figure}[htbp]
\begin{center}
\includegraphics[width=8.9cm]{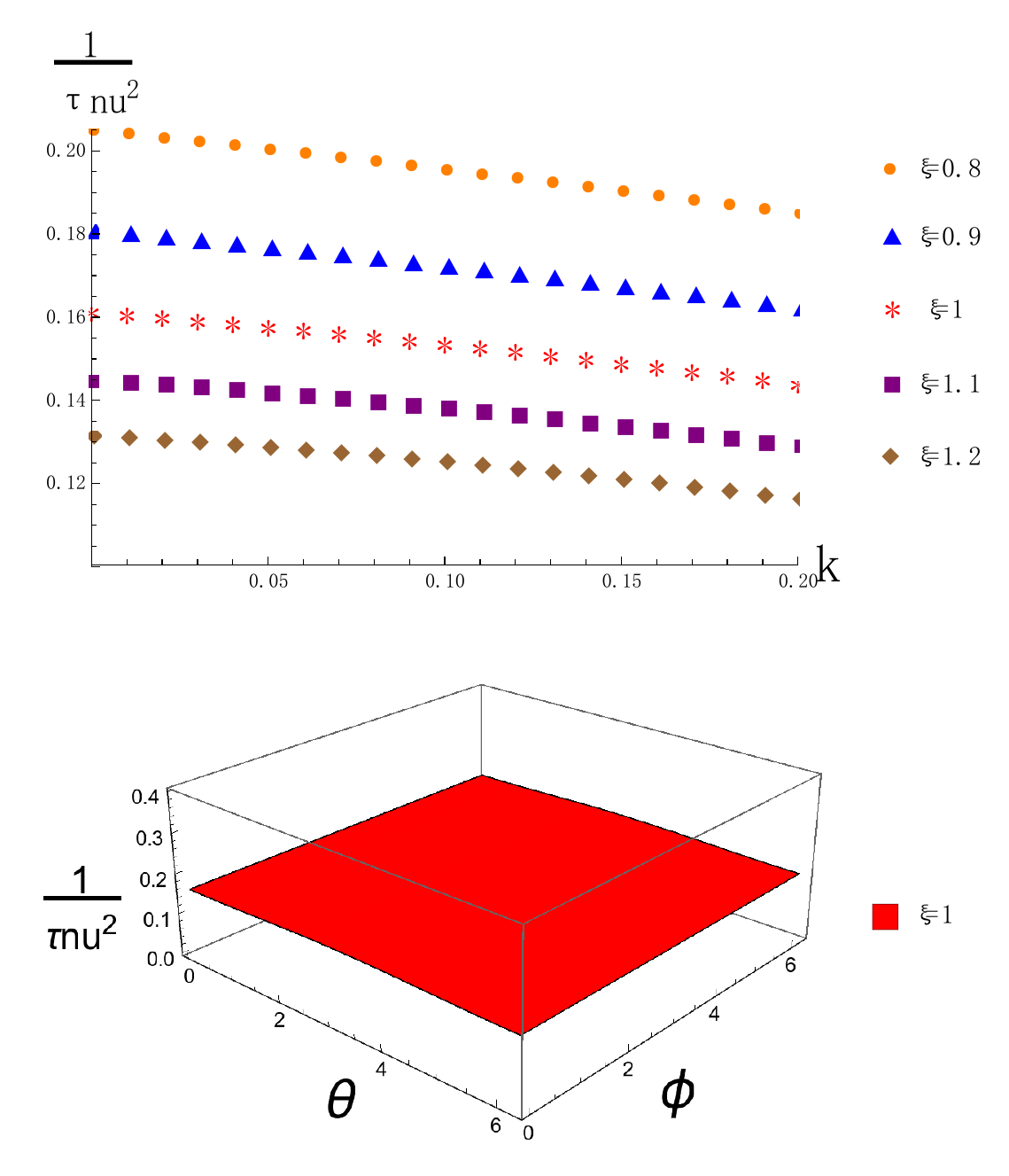}
\end{center}
\caption{$k$-dependence (a) and $angle$-dependence of scattering rate with Yukawa potential for $\xi=0.8-1.2$. From (a) and (b) we can see the $\vec{k}$-dependence is very weak and the $\theta$ and $\phi$-dependence are fitted as $0.165146-0.00436682\cos\theta$.}
\end{figure}

\section{Conclusion and discussion}
In this paper we have studied the zero temperature electronic transport property in nodal line semimetals of small chemical potential with short ranged impurity potential. By studying the Hamiltonian in Eq.(1), we calculate the Drude conductivity analytically. Then we consider the effect of quantum interference, which we find to be negative in both $x$ and $z$ direction, corresponding to the weak localization of electrons. Under the weak magnetic field, the quantum interference corrections $\delta\sigma_{xx}^{qi}$ and $
\delta\sigma_{zz}^{qi}$ behave differently, corresponding to different behaviors of magnetoconductivity. For perpendicular magnetic field, the magnetic field will destroy the weak localization. In the limit of $l_\phi\gg l_B\gg l_z$, the magnetoconductivity $\delta\sigma^{qi}(B)\propto\sqrt{B}$. And in the limit of $l_B\gg l_\phi$ and $l_B\gg\l_z$, it is proportional to $B^2$. For parallel magnetic field, due to the existence of the $\pi$ Berry phase, the magntoconductivity is negatively proportional to $B$, which is consistent with the experiment in Ref.[43]. When the magnetic field parallels to the current, there is no Hall effect, so the resistivity matrix is diagonal. The result $\delta\sigma\propto -B$ means $\delta\rho\propto-\frac{\delta\sigma}{\sigma^2}\propto B$. When the magnetic field does not parallel to the current,the perpendicular component will induce a Hall resistivity $\sigma_{xy}\neq0$. The resistivity is $\rho_{xx}=\frac{\sigma_{xx}}{\sigma_{xx}^2+\sigma_{xy}^2}$, where $\sigma_{xy}$ is the Hall conductivity which is proportional to $B$. We can expand $\rho_{xx}$ around $B=0$, which will give us $\delta\rho_{xx}=c_1B+c_2B^2$.\\

In Ref\cite{Syzranov2017}, the authors considered the transport property in nodal line semimetal with charged impurities\cite{Skinner2013, PhysRevB.93.035138}, which produce a long range scattering potential. They only consider the scattering in a small $k_z$ tube, but here we consider all possible scattering in the whole momentum space. They did not give an analytical calculation of the conductivity and here we give an explicit calculation for $\delta$ potential and we give a numerical result for the screened Coulomb potential.\\

After the completion of this work, we become aware of the paper\cite{PhysRevLett.122.196603}, where the authors discussed two types of impurity potentials, and studied the weak localization and antilocalization effect in detail. We get similar results of $\sigma_{zz}$, and we also calculate the conductivity for parallel magnetic field and find that the magnetoconductivity is proportional to $-B$, which is another feature of the $\pi$ Berry phase and can be used to explain the experiment in Ref. [39].

\section{Acknowledgements}

HY thanks Zhi-Qiang Gao and Kai-Wei Sun for their helpful discussion. FW acknowledges support from 
The National Key Research and Development Program of China (Grand No. 2017YFA0302904)

\appendix
\section{The Drude conductivity and quantum interference}
In this appendix, we give some details about the calculation. We should note that after the coordinate transformation, the Jacobian is given by $\int\frac{d^3k}{(2\pi)^3}\rightarrow\int\;\frac{d\theta}{2\pi}\frac{d\varphi}{2\pi}dkN(k)$ with $N(k)=\frac{k}{2\pi v}$. First we calculate the velocity correction. From Fig. 2b, we have
\beq
\tilde{v}_k^i=v_k^i+\sum_{k^\prime} G_{k^\prime}^RG_{k^\prime}^A\tilde{v}_{k^\prime}^i\mean{ U_{k,k^\prime}U_{k^\prime,k}}_{dis},
\eeq
and $G_k^{R(A)}=\frac{1}{E_F-E_k\pm\frac{\mathbbm{i}}{2\tau}}$. We can calculate the velocity $\tilde{v}^i$ by solving this equation, $\tilde{v}^{x(x)}=v^{x(y)}$ and $\tilde{v}^z=2v^z$. Then we can calculate the Drude conductivity Eq.(5).\\

The vertex $\Gamma_{k_1,k_2}$ satisfies
\beq
\Gamma_{k_1,k_2}=\Gamma_{k_1,k_2}^0+\sum_k\Gamma^0_{k_1,k}\mathcal{G}^e_k\mathcal{G}^h_{q-k}\Gamma_{k,k_2},
\eeq
which can be solved if we assume $\Gamma_{k_1,k_2}=\frac{nu^2}{2}(a+b\cos(\theta_1-\theta_2)+c\sin(\theta_1-\theta_2))$. The result is given by Eq.(8). Then we can calculate the quantum interference correction to the conductivity.
\beq
\sigma_{a1}=\frac{e^2}{2\pi}\sum_q\Gamma_{k,q-k}\sum_kG^R_k\tilde{v}_k^iG_k^RG_{q-k}^R\tilde{v}_{q-k}G^A_{q-k},\nonumber\\
\sigma_{a2}=\frac{e^2}{2\pi}\sum_q\Gamma_{k_1,q-k}\sum_{k,k_1}\tilde{v}_k^i\tilde{v}_{q-k_1}^i\nonumber\\ G_k^RG_{k_1}^RG_{q-k}^RG_{q-k_1}^RG_k^AG_{q-k_1}^A\mean{U_{k,k_1}U_{q-k,q-k_1}}_{dis},
\eeq
which can be calculated in the limit $\omega=0$, $q\rightarrow 0$.

\section{Magnetoconductivity}
In the magnetic field the energy levels is quantized as Landau levels (Eq.(10)). We can use Euler–Maclaurin formula to calculate the summation over Landau index. For perpendicular magnetic field
\beq
\sum_n\frac{1}{(n+1/2)+l_B^2q_z^2}=\ln\frac{l_B^2/l_z^2+1/2+l_B^2q_z^2}{l_B^2/l_\phi^2+1/2+l_B^2q_z^2}+\nonumber\\\frac{1}{2}\left(\frac{1}{l_B^2/l_z^2+1/2+l_B^2q_z^2}+\frac{1}{l_B^2/l_\phi^2+1/2+l_B^2q_z^2}\right).
\eeq
Then we can calculate the integral over $q_z$. In the limit of $l_\phi\gg l_B\gg l_i$, the magnetoconductivity $\delta\sigma_{zz}^{qi}\propto\sqrt{B}$.

For parallel magnetic field
\beq
\sum_n\frac{1}{n\!+\!l_B^2q_z^2}\!=\!\ln\frac{1/l_z^2\!+\!q_z^2}{1/l_\phi^2\!+\!q_z^2}\!+\!\frac{1}{2l_B^2}\left(\frac{1}{1/l_z^2\!+\!q_z^2}\!+\!\frac{1}{1/l_\phi^2\!+\!q_z^2}\right),
\eeq
which is linear as a function of $B$.

\section{The cases without time-reversal and inversion symmetry and with spin orbital coupling}
In sec. II, we have discussed that the Hamiltonian given in Eq.(3) breaks time-reversal and inversion symmetry. We can do the same calculation as in Appendix A and B. We find the velocity correction is given by
\beq
\tilde{v}^{x(y)}=v^{x(y)}\quad\tilde{v}^z=2v^z,
\eeq
but here $v^z=v(\lambda\cos\theta+\sin\theta)$, so all the calculation and the results are almost identical except for the difference in $v^z$.

Considering spin-orbit coupling in Eq.(4), we can do the same calculation. In the limit $\Delta\gg E_F$, we have
\beq
\tilde{v}^{x(y)}=\frac{4}{3}v^{x(y)}\quad\tilde{v}^z=2v^2.
\eeq
In this case the average 
\beq
\mean{U_{k_1,k_2}U_{k_2,k_1}}_{dis}=\mean{U_{k_1,k_2}U_{-k_1,-k_2}}_{dis}\nonumber\\=nu^2\cos^2\frac{\theta_1-\theta_2}{2}\cos^2\frac{\phi_1-\phi_2}{2}.
\eeq
Repeat the calculation in Appendix A and B, we will get the Drude conductivity and the quantum interference vertex
\beq
\sigma_{xx}^{sc}=\sigma_{yy}^{sc}=\frac{e^2N_F\tau}{3},\quad \sigma_{zz}^{sc}=e^2N_F\tau.\\
\Gamma_q=\frac{1}{\frac{1}{4}(q_x^2+q_y^2)\tau^2+\frac{1}{2}q_z^2\tau^2}
\eeq
The results here are qualitatively the same as the results in the main text. 

For the Yukawa potential case $\Gamma_{k_1,k_2}^0=\frac{1}{2}(1+(\cos\theta_1-\theta_2))\frac{1}{(1+(\vec{k}_1-\vec{k}_2)^2\xi^2)^2}$, from which we can calculate $D_1^\prime$ and $D_2^\prime$ by expanding in $\Gamma_0$ in terms of $\vec{k}$ and keeping the constant terms only.
\bibliographystyle{apsrev4-1}
\bibliography{conductivityinNL.bib}

\end{document}